\documentclass{aa}
\usepackage{graphicx}

\begin{document}

\sloppypar

   \title{The spectrum of the cosmic X-ray background observed by RTXE/PCA}

   \author{M. Revnivtsev\inst{1,2}, M. Gilfanov \inst{1,2}, R. Sunyaev \inst{1,2}, K.Jahoda \inst{3}, C.Markwardt \inst{3}}

   \offprints{mikej@mpa-garching.mpg.de}

   \institute{Max-Planck-Institute f\"ur Astrophysik,
              Karl-Schwarzschild-Str. 1, D-85740 Garching bei M\"unchen,
              Germany,
        \and   
              Space Research Institute, Russian Academy of Sciences,
              Profsoyuznaya 84/32, 117810 Moscow, Russia
	\and 
		Laboratory for High Energy Astrophysics, Code 662, Goddard Space Flight Center, Greenbelt, MD 20771, USA
            }
  \date{}

        \authorrunning{Revnivtsev et al.}
        \titlerunning{}
 
   \abstract{We have 
analyzed a large set of RXTE/PCA 
scanning and slewing observations performed between April 1996 and March 1999. 
We obtained the 3-20 keV spectrum of the cosmic X-ray background (CXB) by subtracting
Earth-occulted observations from observations of the X-ray sky at high galactic
latitude and far away from sources.
The sky coverage is approximately 
$\sim22.6\cdot 10^{3}$ deg$^2$. The PCA spectrum of CXB in 3-20 keV  energy band 
is adequately approximated  by a single power law with photon index 
$\Gamma\sim 1.4$ and normalization at 1 keV $\sim9.5$ phot/s/cm$^2$/keV/sr.
Instrumental background uncertainty precludes accurate RXTE/PCA
measurements of the spectrum of cosmic X-ray background 
at energies above 15 keV and therefore
we can not detect the high energy cutoff observed by HEAO-1 A2 experiment. 
Deep observations of the 6 high latitude points used to model the PCA background 
provide a coarse measure of
the spatial 
variation of the CXB. 
The CXB variations are consistent with a fixed spectral shape and variable normalization
characterized by a fractional rms amplitude of $\sim$7\% on angular scales of $\sim$1 square deg.
   }

   \maketitle

%

\section{Introduction}

Many efforts over the last few decades have contributed to an understanding of the
origin of the cosmic X-ray  background (CXB) in the 2-10 keV band
(e.g. \cite{boldt87}, \cite{hasinger91}, \cite{fabian92}, \cite{mushotzky00}, 
\cite{giacconi02}, \cite{brandt03}). 
Most, if not all, of the CXB emission is explained by the 
superposition of point sources (AGNs) distributed over the Universe 
(see e.g. \cite{rees80}, \cite{giacconi87}, \cite{setti89}, \cite{giacconi02}, \cite{moretti03}).

Accurate measurements of the X-ray spectrum of the CXB 
 were obtained from large solid angle measurements with collimated 
spectrometers aboard the HEAO-1 observatory.  The HEAO-1 A2 experiment
was designed specifically for this problem, with special care being taken
to separate the signal from cosmic and instrumental backgrounds.
The A2 proportional counter measurements from $\sim 3 - 60$ keV (\cite{marshall80}) are
extended to 1 MeV with the scintillators of the A4 experiment (\cite{gruber92}).
Numerous other measurements have been made with imaging
telescopes - EINSTEIN, ROSAT, ASCA, BeppoSAX, CHANDRA, and XMM 
(e.g. \cite{wu91}, \cite{gendreau95}, \cite{chen96}, \cite{miyaji98}, \cite{vecchi99}, \cite{mushotzky00}, \cite{lumb02}). 
The HEAO-1 A2 
measurements were made over a large solid angle with an instrument designed to ensure a precise
instrumental background subtraction. 
The imaging experiments measured the CXB 
over a much smaller solid angle
and could be
subject to cosmic variance (see e.g. \cite{barcons92}, \cite{barcons00}). 
The measurements made by X-ray telescopes all yield
an absolute normalization significantly larger than that 
of HEAO-1 A2, a result difficult to explain by cosmic variance alone (e.g. \cite{barcons00})

The Proportional Counter Array (PCA) aboard Rossi X-ray Timing Explorer (RXTE) provides an
opportunity to perform a new and
independent measurement of the CXB
spectrum based on nearly all sky data --
the first such measurement since the HEAO-1 observations. 

\section{Data analysis and results}
\begin{figure}
\includegraphics[width=\columnwidth]{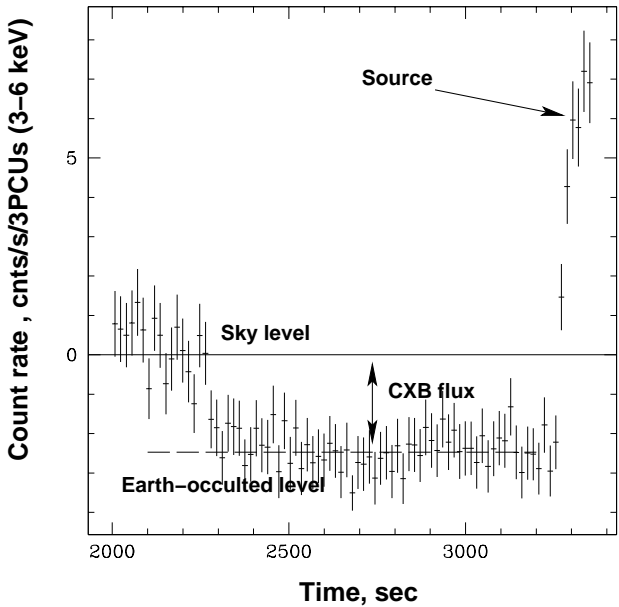}
\caption{Typical PCA background subtracted lightcurve during  slew.
The clear difference between the ``sky'' level and ``Earth'' level measures the CXB flux. 
The increase in countrate at the end of the lightcurve is caused by an 
X-ray source becoming un-occulted. \label{lcurve_demo}}
\end{figure}

\begin{figure*}[t]
\includegraphics[bb=40 240 560 540,clip,width=2\columnwidth]{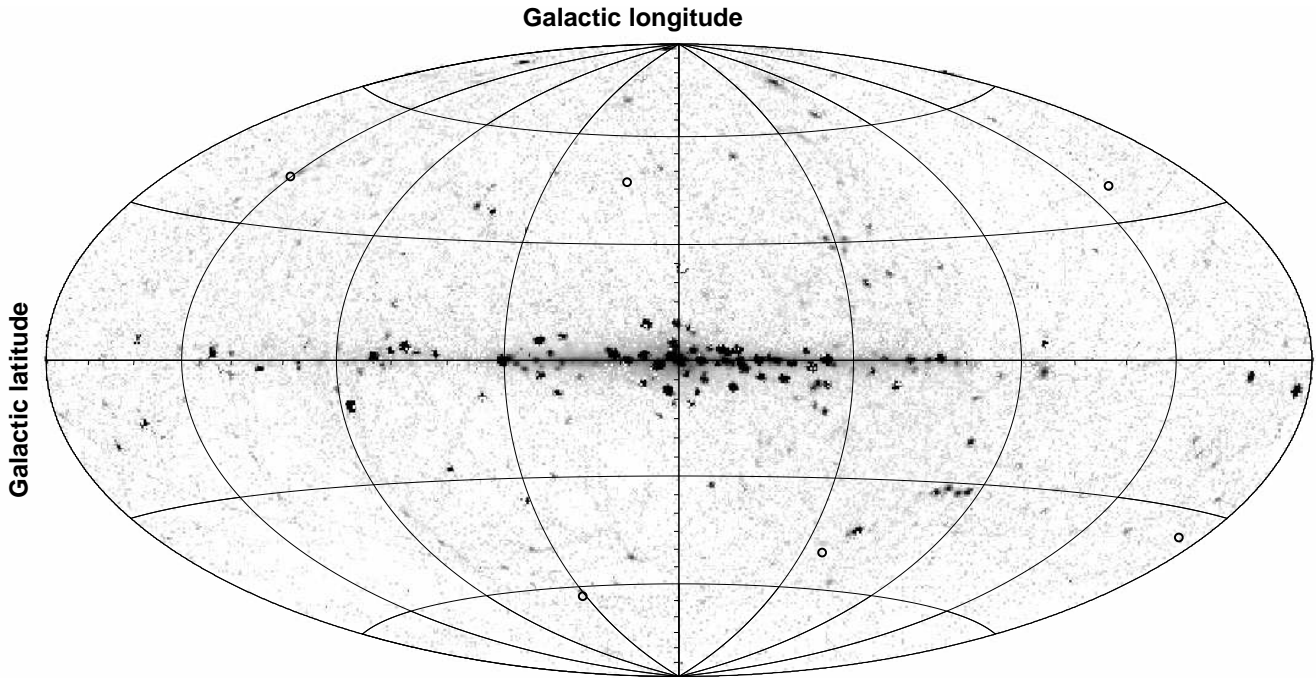}
\caption{Map of the sky, reconstructed from slew observations of RXTE/PCA. 
For this map we used data from layer 1 of PCUs 0,1,2 , 3-20 keV energy band.
The grid is separated by 45$^\circ$ in $l$ and 30$^\circ$ in $b$. Open circles indicate
the regions used to construct the PCA background model.
\label{skymap}}
\end{figure*}

\subsection{Data and software}

The Rossi X-ray Timing Explorer (\cite{rxte}) carries three instruments including
the X-ray spectrometer: the Proportional Counter Array (PCA). It consists
of 5 independent Proportional Counter Units (PCUs) which are sensitive to
photons in 2-60 keV energy range.  For Crab-like spectra,
88\% of the detected
counts are below 10 keV.
Due to its high 
effective area ($\sim$6400 cm$^2$ at 6-7 keV), relatively precise modeling
of the instrumental background, and low deadtime, the PCA can reliably measure spectra
for sources with flux greater than 1 mCrab, which is about the flux of the CXB
integrated over the 1 degree beam of the PCA.

The RXTE is capable of fast slews (6 deg/min);  typical operations 
include 1-2 slews per orbit.  Although operations are planned to slew during the South Atlantic
Anomaly or periods when targets are occulted to the greatest extent possible,
a substantial amount of blank sky and dark earth occulted data is obtained.
In addition, several Guest observer programs requested scanning observations
over moderate areas of the sky.  The slewing and scanning data can be used
to construct maps of the sky with the  $\sim 1 \deg$ resolution of the
PCA collimator.
(e.g. \cite{lens}). Scanning RXTE/PCA 
observations are very useful for the 
localization of newly discovered sources (\cite{craig00})
and the study of extended structures on the X-ray 
sky, especially at relatively high energies (10-20 keV), where only
limited amount of data exists (see e.g. \cite{valinia96}, \cite{mikej03}).   

In our study we used 
RXTE/PCA data taken during reorientations (slews or scans)
from April 16, 1996 through March 22, 1999. This time period
was chosen to stay within a single high voltage epoch of PCA.
The total number of
observations is approximately 17,600 with $\sim$8.5 Msec of exposure.
These data contain both clean-sky observations and Earth occulted observations.
Clean-sky data provide the
cosmic X-ray background signal, while Earth-occulted data
provide information about the PCA instrument background.

The first attempts to model the PCA background used earth looking data
as an estimate of the instrument background.  No separation between
dark and sunlit earth was made.  The estimated CXB spectrum, using
this background estimator, deviated from a power law with index $\sim -1.4$
at $\sim 15 keV$, and also some soft component appeared, an effects that were
attributed to reflection from the bright earth.  
Experience with BBXRT and ASCA (albeit at lower energies)
suggests that the sunlit earth is more than an order of magnitude brighter
than the dark earth.  As we have taken care to include only dark earth
data, and the statistically significant signal extends only to 15-20 keV, we
are confident that the earth albedo is effectively zero for this experiment.
At these low energies the Earth atmosphere, consisting of nitrogen and oxygen,
is a very effective absorber, however at energies higher than 10-15 keV
effect of reflection from the atmosphere plays an important role 
(see e.g. \cite{pendleton92}) and should be taken into account.

Our analysis assumes that the dark Earth emits
essentially zero flux in X-rays. In reality the emission of the dark Earth
at higher photon energies is modified by the reflection of cosmic X-ray 
background and radiation from brightest X-ray sources. However
below we would assume that the influence of this effect is small
in the spectral band of our interest.

Data reduction was done using standard tools of LHEASOFT 5.2 package.
We analyzed data from PCA detectors PCU 0, 1 and 2, which have largest
exposure times. Noisy parts of data were filtered out by applying the
selection criteria $ELECTRON\_0,1,2<0.1$.
All results were corrected for the deadtime 
(http://legacy.gsfc.nasa.gov/docs/xte/recipes/ pca\_deadtime.html). 

The effective field of view of the PCA is $\Omega=2.97\cdot 10^{-4}$ sr.  
This value
is derived by fitting scans over the Crab nebula to a model which convolves the
response of a perfect hexagonal collimator (8 inches high with a 1/8 inch flat to flat
opening) with a Gaussian (FWHM 6 arcmin, \cite{jahoda96}).  This model is appropriate for many independent and nearly co-aligned
hexagonal collimators;  the width of the Gaussian characterizes the average misalignment.
While construction of the PCA collimators from corrugated sheets soldered together
causes strong correlations between nearby collimator cells, the model works adequately to
describe the ensemble of $\sim 20,000$ collimator cells per PCU.

To check the overall normalization (i.e. net area) of the PCA we analyzed
the Crab monitoring observations obtained approximately every two weeks over the
entire period during which CXB data was collected.
We selected data from the same anodes as described
above (layer 1, PCUs 0,1 and 2). With standard response matrix, produced
by FTOOLS 5.2 package the photon index of the Crab spectrum in the  3-20 keV
energy band was measured to be $\Gamma=2.09\pm0.01$ with a normalization of
$N=11.6\pm0.4$ phot/s/cm$^2$/keV (the value of neutral absorption column
was fixed at $N_H=2\cdot 10^{21}$ cm$^{-2}$;  the results in the PCA band are insensitive to this).

This fit predicts a Crab nebula flux of $ 2.66 \cdot 10^{-8}$ erg/s/cm$^2$ (2-10 keV)
which is high compared to the ''conventional'' value.
Zombeck (1990) reports the Crab spectrum as presented by Seward (1978) to be
 $dN/dE = 10 E^{-2.05}$
phot/s/cm$^2$/keV, which gives a 2-10 keV flux of $ 2.39 \cdot 10^{-8}$ erg/s/cm$^2$.
To our knowledge, more recent X-ray experiments have not measured this normalization
directly, using instead this value as a standard candle. To put our measurements
on this scale, we correct our measured fluxes downward by a factor of $1.11$;  this
is equivalent to increasing the estimated geometric area of the PCA.  
\footnote{We (the GSFC co-authors) expect that the FTOOLS 5.3 package, anticipated for
fall 2003 release, will increase the geometric area parameterization in the xpcaarf tool,
which will effectively reduce derived fluxes by this factor;  given the good agreement
in the spectral index between PCA and the other measurements, this boot-strap approach
represents the best available calibration of the total area.}
A multi-mission attempt to use type I X-ray bursts as standard candles reached
a similar conclusion (\cite{kuulkers03}) using an earlier version of the PCA response matrix.

Deviations between the modeled and measured Crab
spectrum do not exceed $\sim$1\%. In all subsequent analysis
we cite only statistical uncertainties unless 
otherwise noted.  The quoted uncertainties are
consistent with the statistical distributions of measured quantities.

We used the faint source (''L7\_240'') CM background model
(http://heasarc.gsfc.nasa.gov/docs/xte/ recipes/pcabackest.html). 
The background model includes by design both the cosmic and instrumental
background, so the background subtracted rate for ''blank sky'' observations should be approximately
zero.  The background model is constructed from observations of six
different blank sky points;  the net background rate is slightly different
for the six points due to spatial fluctuations of the CXB.

Combination of 
Earth-occulted 
observations with sky observations provides 
the spectrum of the CXB. The Earth is treated 
as a shutter in the front of the PCA.  Fig. \ref{lcurve_demo}
illustrates this point.
The ``sky'' rate
is approximately zero; the decrement between sky and occulted data is 
just the CXB flux.

\subsection{Map of the sky}

In order to obtain the spectrum of the CXB,
we need to avoid the contamination from bright galactic and extragalactic sources
present in the data. For this purpose we have constructed the map of the sky using the
same Standard2 data mode from which we collect the CXB spectrum. 

The Standard2 data mode of the PCA is present in all observations,
maintains the maximum useful energy resolution, separates data by anode,
and provides 16 second time resolution.
 The background subtracted flux, measured by
each PCU in 3-20 keV energy band during each 16 second time bin 
was ascribed to the point of the sky where the optical axis 
of the RXTE/PCA was pointed at the middle of the time bin. 
Angular resolution is limited by the size of the PCA beam ($\sim 1 ^\circ$ FWHM) 
and the movement of the optical axis during each 16 sec interval.
The typical velocity 
of the RXTE optical axis on the sky -- $\sim$0.1 deg/sec -- limits 
the spatial resolution along the slew
direction $\sim 1-1.5^\circ$. 
The RXTE/PCA collimator field of view is $\sim1 ^\circ$ (FWHM). Therefore 
the Standard2 data provides a skymap with $\sim1-1.5^\circ$
resolution. The map is presented in Fig.\ref{skymap}.

Dark $\sim1^\circ$ circles represent point sources, and the dark 
bar along the Galactic plane is caused by the Galactic ridge diffuse emission.
Sco X-1 is not seen on the picture, because its strong X-ray flux leads
to violation of our criteria of filtering the ``bad'' data.

The $3\sigma$ sensitivity of the obtained map to the point sources is 
approximately at the level of $\sim10^{-11}$  erg/s/cm$^2$.
At the present time this all sky map is most sensitive in 
the energy band 3-20 keV. In addition to that data of RXTE observations
in the period 1999-2002 can provide us approximately 2 times more 
effective exposure of the sky resulting
in the all-sky map limiting sensitivity $\sim 0.5\cdot 10^{-11}$ 
erg/s/cm$^2$. Analysis of the point sources is beyond the scope 
of this paper. We plan to present such analysis as a separate work.
 
Data within  1.5$^\circ$ of all detected sources was masked.
We have excluded the
data obtained at low Galactic latitudes ($|b|<20^\circ$) in order to
avoid the influence of the Galactic ridge diffuse emission and weak 
Galactic X-ray sources. Regions of $10^{\circ}$ around LMC and SMC 
were also excluded. 

 Analysis of the latitude profiles of the Galactic 
ridge emission (\cite{iwan82}, \cite{valinia96}, Revnivtsev et al. 
2003) shows that its contribution to the detected X-ray flux at latitudes 
$|b|>20^\circ$ is less than approximately $10^{-2}$ cnts/s/PCU, and therefore
negligible for our study (CXB has $\sim$2 cnts/s/PCU).

After exclusion of all mentioned regions we have approximately 1.7 Msec 
of data, covering $\sim$55\% of the sky ($\sim22.6\cdot 10^3$ deg$^2$) with 
non-zero exposure.  Our sensitivity at higher energies is limited by the
statistics of the background removal;  our dark earth spectrum contains
only 25 ksec of data.

\subsection{Spectrum of CXB }

The unfolded spectrum of CXB is presented in Fig.\ref{spectrum}. 
The spectrum is well approximated by a single power law with a photon index
$\Gamma=1.42\pm0.02$ and normalization $N=9.8\pm0.3$ phot/s/cm$^2$/keV/sr.
If the slope is fixed at $\Gamma=1.4$, the normalization
is $N=9.5\pm0.3$ phot/s/cm$^2$/keV/sr. 
The observed flux of CXB in 3-10 keV energy band is 
$F_{\rm 3-10 keV}=(5.17 \pm0.05) \cdot 10^{-8}$ erg/s/cm$^2$/sr. 
The normalizations and flux are reported after the downward correction of 1.11
described above.

The study of CXB emission with PCA is limited by the accuracy of the
instrumental background subtraction above 10 keV. Unmodelled variation in
the PCA background averages 0.027 ct/s/PCU in the 2-10 keV band and
0.013 ct/s/PCU in the 10-20 keV band (these figures are layer 1 only 
\cite{craig02} 
and thus relevant to the data discussed here).

These errors were added in quadrature to
the statistical errors of the measured count rates.
The net statistical and systematic uncertainties of measured CXB spectrum
are shown by shaded area in Fig.\ref{spectrum}. 

\subsection{Cosmic variance}

Measurements of the CXB are also affected by spatial fluctuations, or cosmic variance, over the sky
(see e.g. \cite{barcons00}). 
Our CXB spectrum, which averages over a huge solid angle, is expected to
represent a well defined average, although the background model, which is
based on observations of only 6 distinct deep exposure points, may be affected.
Other instruments that measure the CXB spectrum
over smaller solid angles may be directly affected by this variance.

\begin{figure}
\includegraphics[width=\columnwidth,bb=28 182 566 715,clip]{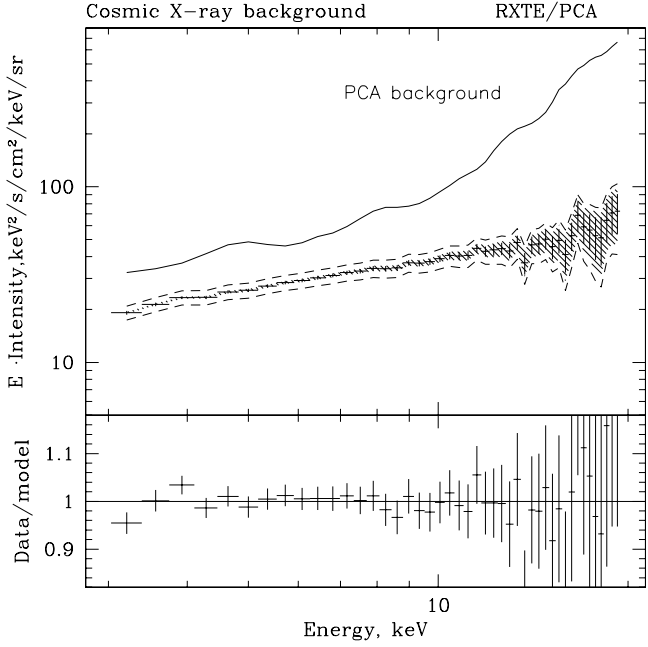}
\caption{Unfolded flux spectrum of CXB obtained by RXTE/PCA. Shaded area around the spectrum
represents the amplitude of systematic uncertainties in the
background subtraction. The spectrum of the PCA background (internal + CXB) 
 is shown by the 
solid line.  Dashed lines represents root-mean-square amplitude of 
variations of normalization of CXB (cosmic variance) measured over different sky areas with effective solid angle $\sim$1 sq.deg. (see text). 
Lower panel shows the ratio of observed data to used model.\label{spectrum}}
\end{figure}

The data used to create the background model also allows a limited measurement
of the fluctuations on the scale of the PCA beam of $\sim1^\circ$.
We analyzed data from the 6 ``background'' points (marked with open
circles in Fig.\ref{skymap}).
After subtraction of the PCA instrument plus average sky background, 
the spectra of 
these sky areas are significantly different.  The individual spectra
 can be described with
a common power-law index but differing normalization.
The fractional root-mean-square amplitude of this variation in normalization
is $7\pm1$\%. 
The dashed lines in Fig. \ref{spectrum} show the $\pm 1 \sigma$ range of 
normalizations that 
would be measured over 1 square degree solid angles given this variation.

Field to field variation of the CXB is due to a combination of Poisson
noise associated with different populations of sources drawn from the
log N-log S distribution and to large scale structure.  For the 1 degree field
of view of the PCA, the Poisson variance is expected to be $\sim 4\%$ (\cite{barcons98a}; \cite{barcons98b}).

This suggests that the PCA background fields have measured large scale structure in the 
CXB, an interpretation consistent with {\it Chandra} measurements of the CXB
fluctuations of  $\sim 25-30\%$ on scales of $\sim0.07$ sq.deg (\cite{yang03}), after
scaling by the square root of the solid angle.

\section{Conclusion}

The large amount of slew data from the RXTE/PCA instrument
allows us to study the spectrum and intensity of cosmic X-ray background
averaged over a large solid angle.

After excluding areas around
bright sources, Galactic plane region ($|b|<20^\circ$) and regions
of Large and Small Magellanic Clouds our data
covers approximately $22.6\cdot 10^3$ deg$^2$ of the sky. 
This data set measures,
by definition, the average properties of the CXB.

The spectrum of the CXB in the 3-20 keV energy band obtained from RXTE/PCA slew data is 
well approximated by a power law in the form $dN(E)/dE=N E^{-\Gamma}$ 
with photon index $\Gamma=1.42\pm0.02$ and normalization $N=9.8\pm0.3$ phot/s/cm$^2$/keV/sr.
Relatively large systematic uncertainties of the PCA instrumental background at E$>$15-20
  keV, and the decreasing ratio of CXB to instrument background
 did not allow us to study the spectrum above  $\sim$20 keV, and the high energy cut-off 
detected by HEAO-1 A2.

\begin{figure}
\includegraphics[width=\columnwidth,bb=28 182 566 715,clip]{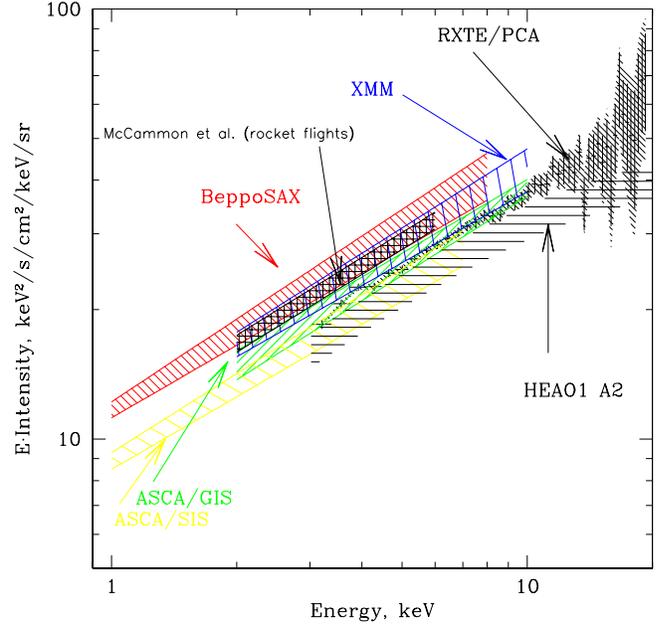}
\caption{Spectrum of CXB obtained by different instruments \label{dif_spectra}}
\end{figure}

\begin{figure}
\includegraphics[width=\columnwidth,bb=28 182 566 715,clip]{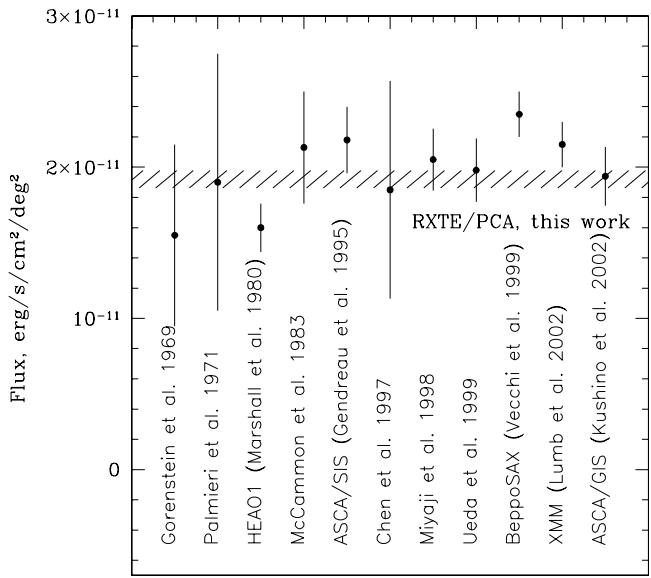}
\caption{Comparison of level of CXB obtained with RXTE/PCA with previous measurements
\label{dif_measurements}}
\end{figure}

The slope of CXB spectrum obtained by RXTE/PCA agrees well with that obtained
by other observatories. The normalization value is slightly higher, 
but marginally compatible with, HEAO-1 A2 (\cite{marshall80}), obtained over a
similarly large
solid angle of the sky. The measurements of CXB normalization by different
X-ray instruments give different
variable values (Fig. \ref{dif_spectra}, \ref{dif_measurements}), however,  the weighted average 
of the imaging measurements gives a value that is inconsistent with HEAO-1 A2  
(\cite{barcons00}) and consistent with ours.

Our measurement relies on a scaling of the absolute area of the PCA to match
the canonical value of the flux from the Crab nebula;  as many 
experiments have used the Crab as a standard candle and as our 
measurement of the spectral shape is in good
agreement, this should add no more than a few per cent uncertainty.
Our result is marginally compatible both with
the result from the Xenon-filled, collimated proportional counters 
of HEAO-1 A2 and with
the results from imaging instruments. However, observed subtle discrepancies
between these measurements suggest that there are remaining systematic
issues in the calibration of the effective area and/or solid angle in 
either the collimated or imaging experiments.

\begin{acknowledgements}
Authors are grateful to R. Mushotzky for valuable comments and suggestions.
This research has made use of data obtained through the High Energy
Astrophysics Science Archive Research Center Online Service,
provided by the NASA/Goddard Space Flight Center.
\end{acknowledgements}

\end{document}